# Towards Modeling Cross-Domain Network Slices for 5G

Rami Akrem Addad[1], Tarik Taleb[1], Miloud Bagaa[1], Diego Leonel Cadette Dutra[2]
and Hannu Flinck[3]
[1] Aalto University, Espoo, Finland
[2] Federal University of Rio de Janeiro, Rio de Janeiro, Brazil
[3] Nokia Bell Labs, Espoo, Finland

*Abstract*—Network Slicing (NS) is expected to be a key functionality of the upcoming 5G systems. Coupled with Software Defined Networking (SDN) and Network Function Virtualization (NFV), NS will enable a flexible deployment of Network Functions belonging to multiple Service Function Chains (SFC) over a shared infrastructure. To address the complexities that arise from this new environment, we formulate a MILP optimization model that enables a cost-optimal deployment of network slices, allowing a Mobile Network Operator to efficiently allocate the underlying layer resources according to the users' requirements. For each network slice, the proposed solution guarantees the required delay and the bandwidth, while efficiently handling the usage of underlying nodes, which leads to reduced cost. The obtained results show the efficiency of the proposed solution in terms of cost and execution time for small-scale networks, while it shows an interesting behavior in the large-scale topologies.

## I. Introduction

Unlike the previous generation of mobile networks, 5G systems are expected to rely on both the advancement of physical infrastructures represented by the introduction of Millimeter waves, massive MIMO, full duplex, beam-forming, and small cells; as well as the emergence of Software Defined Networking (SDN) and Network Function Virtualization (NFV) [1], [2]. By introducing the logical infrastructure abstraction, the 5G mobile networks will completely transform modern network infrastructures as SDN and NFV paradigms represent the key enabler technologies towards the softwarized networks. Network Softwarization represents one of the main keys for enabling the most suitable 5G's use cases, by reducing both the Capital Expenditures (CAPEX) and the Operating Expenditure (OPEX), while allowing an easy deployment schema [3]–[6]. Network Softwarization can enable high-performance improvements by offering the flexibility and modularity that are required to create multiple network slices. These facilities offered by the softwarized networks cede place to a new concept dubbed Network Slicing [7], [8].

The network slicing paradigm will play a major role to efficiently implement different 5G's use cases related to distinct verticals with divergent requirements over the same network infrastructure such as enhanced Mobile Broadband (eMBB), Ultra-Reliable and Low Latency Communications (uRLLC) and massive Machine Type Communication (mMTC) [9]. However, with the introduction of the micro-segmentation approach which requires almost all the flows to pass through several Network Functions (NFs), the introduction of a mechanism that takes over becomes a must.

Service Function Chaining (SFC) is foreseen to be a solution that will dynamically steer the network traffic and flows across multiple physical and logical infrastructures [10], [11].

Given the circumstances, it is clear that the new trend consists of instantiating network slices that contain one or more SFCs [12]. Each SFC forms a set of NFs running inside either a logical node or a physical node. In order to enable the Slice-SFC approach, many NFs may require being traversed in a certain strict order, leveraging on the flexibility of NFV, Mobile Network Operators (MNOs) can deploy any particular slice type honoring its real-time requirements. Moreover, a NF can run either on top of virtual machines or containers. This flexible management can lead to a huge number of active nodes in the network infrastructure that are scarcely used which leads to an inefficient network slicing deployment. To address this issue, this paper introduces an optimization solution for achieving an efficient network slicing deployment while keeping the cost minimized.

The remainder of this paper is organized as follows. Section II summarizes the fundamental background topics and related research works. Section III describes the proposed architecture and our network model. Section IV illustrates the problem formulation and describes our proposed framework solution. Section V presents the performance evaluation and our results analysis. Finally, Section VI concludes the paper.

## II. Related work

Moens and Turck [13] present and solve a VNF placement problem. Their evaluation shows that the developed algorithm finishes 16 times faster for a small service provider than the previous solutions. In [14], Ko et al. present an optimal placement of network functions in an SFC context. The problem was solved by considering the latency required to place the service function in a given SFC. Their model is an Integer Non-Linear Programming (INLP) problem based on the latency requirements.

Song et al. [15] treat the problem of obtaining an optimal placement of network functions in the operators' networks.

They formulated an Integer Linear Programming (ILP) problem to demonstrate the trade-off between the network cost and the computing resources cost. Their network cost is mainly the bandwidth of the links between the network functions, leaving out the link delay, and the resources cost are the CPU consumed by these network functions. Jiao et al. [16] presented a solution to a similar environment as in [15], changing the problem's objective. In this paper, the authors tried to maximize the traffic throughput under the constraint of the end-to-end latency in a given service function chain to obtain an optimal placement.

In [17], the authors evoke the problem of coordinated NFV Resource Allocation (NFV-RA). The authors try to minimize an overall cost which is constituted of the link cost (bandwidth, latency), the CAPEX cost and the OPEX cost by using an improvised modeling called Homogeneous Link Mapping (HLM). Wen et al. [18] proposed a Network Function Consolidation (NFC) modeling, followed by an ILP formulation, that they solved using a greedy-based heuristic solution. The authors tried to minimize the number of VNFs deployed in the network by allowing several network functions to be hosted in a limited number of VNFs.

### III. PROPOSED ARCHITECTURE & NETWORK MODEL

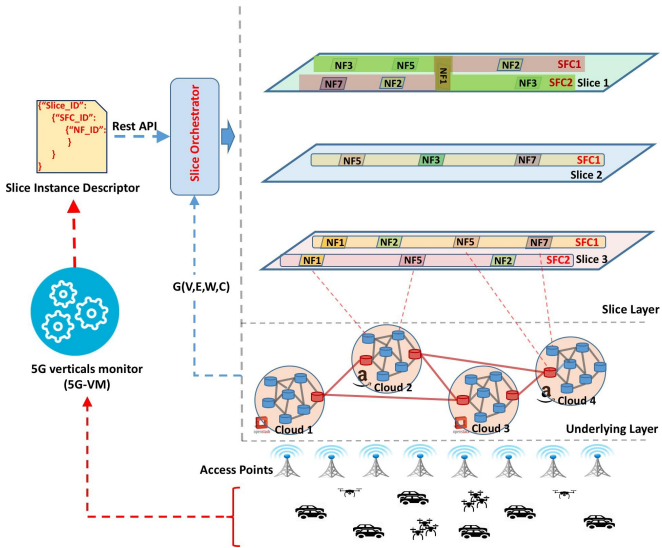

Fig. 1: Global architecture of the proposed solution.

Fig. 1 depicts the architecture envisioned in this paper. It is divided into the underlying layer and the slice layer. The underlying layer is composed of a set of nodes, virtual or physical, and routers. In this layer, the nodes are grouped into a set of computational resources that communicate between themselves through the physical network, and the routers are used as connectors between different computational resources. The slice layer, which runs on top of the underlying layer, consists of a set of slices that are dedicated to different services, e.g., health-care and connected cars. The traffic within each slice is routed using SFC. Each slice is formed by ingress, egress nodes, and a set of intermediate nodes. At the reception of different packets at the ingress node, which is also called classifier, the SFC of those packets would be identified, and then the traffic would be forwarded according to that specified SFC. For instance, in the case of connected car management, a slice can be managed by two different SFCs. While the first SFC could be dedicated to the monitoring and getting different measurements, the second SFC could be used for applying different management actions.

Let $G(V, E, W, C)$ be a weighted graph that represents the underlying layer. Each node represents a virtual or physical instance in a data center. $V = H \cup U$, where $U$ presents the set of nodes deployed in each data center and $H$ denotes the set of connector nodes that connect different data centers. A node in $U$ can be a server, a virtual instance (containers or virtual machines) or a router that forwards the traffic. Meanwhile, the nodes in $H$ form a wide area network (WAN) that interconnects different data centers. Each vertex $v \in V$ consists of an ordered list, where each element in that list describes the number of resources on that node, such as CPU, Memory, and Disk. For instance, a vertex $v$ can be presented as follows $(CPU, RAM, Disk, I/O)$. From another side, $E$ represents all physical links relaying between the nodes $V$. Formally $(u, v) \in E$ if there is a direct link between vertices $u$ and $v$. We also define $E(u, v)$ that shows the relations between two vertices $u, v \in V$. $E(u, v) = 1$ if there is a physical link between $u$ and $v$, otherwise $E(u, v) = 0$. $W$ represents the weight of every link in the physical network $G$ in the form of another ordered list that consists of the bandwidth and the latency $(Bw, L)$. Due to the limited capacity in WAN connection, usually, the links' capacities between $H$ are too low compared to the ones between $U$. Let $W_B(u, v)$ and $W_L(u, v)$ denote the available bandwidth and end-to-end latency between nodes $u$ and $v$. Formally, $W_B(u, v) \leq M \times E(u, v)$ and $W_L(u, v) \leq M \times E(u, v)$, such as $M$ is a big number ($M + \infty$). We denote by $C$ the characteristics of different vertices including the level of security and IaaS of different Nodes. For each node $v \in U$, we denote by $v(C_S)$ and $v(C_I)$ the security level and IaaS of node $v$. The proposed solution can easily consider more complicated characteristics by updating $C$.

Each user starts by sending a slice creation request in order to create a given slice and consume a certain service, such as an end-user represented by a connected car which asks for autonomous driving assistance. The 5G verticals monitor (**5G-VM**) will be in charge of gathering the information about the services running on different users and devices, such as the amount of bandwidth and end-to-end delay. **5G-VM** periodically monitors and detects the changes occurring in the network, including users' and devices' demands and/or their mobility, which can affect the service level agreement (SLA), then it will trigger the Slice Orchestrator (**SO**) for creating and/or rescheduling the different slices, as well as transfer the different monitoring information to the **SO** as a Slice Instance

Descriptor (**SID**). By leveraging the **5G-VM**, the end user submits as a **SID** all the set of necessary specifications for the creation of a slice $s \in S$ to the **SO**. The specifications are represented by the totality of service function chaining $f_i \in F_s$, with $s \in S$, needed for the establishment of a network slice, the ingress node (classifier) and the egress node. While the ingress node classifies the incoming packets based on the pre-defined network policy traffic for the available set of SFCs $F_s$ in the slice $s \in S$, the egress node forwards the processed packets to the outside of the SFC domain (an output node). It is worth noting that the ingress node and the egress node are not included in the set of NF types available in the network.

On another side, the SFCs are composed of a set of NFs connected through virtual links. Each NF requires a certain *CPU*, *RAM* and a set of authorized nodes $y_j$ for the deployment of a given NF. It shall be noted that an end user can impose certain affinity constraints in order to deploy his NFs, e.g., a user can ask for only the public IaaS, which means that only the data centers responsible for hosting the public IaaS will be taken into consideration. Depending on service characteristics, SFCs have different bandwidth and latency requirements. In addition to the bandwidth and the latency between each two NF nodes, the SFCs have the global bandwidth and latency requirements denoted by $l_f$ and $w_f$ that must be satisfied.

After specifying the total requirements for the slice creation, the **5G-VM** transfers the requests to the **SO** in a **SID** format. The **SO** will then take, as an additional input, the physical layer's available resources transmitted from the physical layer. The **SO** will verify for each request from the **5G-VM** the requirements for the slice creation. For instance, if a given NF requires 5G of vRAM, 3 vCPU and a set of authorized IaaS $y_j\{2, 5\}7$, the **SO** will use the available resources from the physical layer to place the NF in the right place in concordance with the other NFs that belong to the same SFC in order to satisfy also the requirements of the connectivity (bandwidth and latency) and finally create the desired slice.

## IV. CROSS-DOMAIN COST AWARE NETWORK SLICING DEPLOYMENT

As mentioned before, each slice $s \in S$ has a set of SFCs $F_s$. Each SFC $f_i \in F_s$ consists of a set of NFs that are interconnected, whereby each NF has one predecessor and one successor except the first and the last NFs. While the first NF has only one successor, which is the second NF in the SFC, the last one has only one predecessor. We denote by $\Psi^s_i$ the set of NFs in the SFC $f_i$. We also denote by $\Psi^s_{i,j}$ the $j^{th}$ NF in SFC $f_i \in F_s$ at the slice $s \in S$. Let $\Gamma$ denote the number of NFs in the network. Formally, $\Gamma$ can be defined as follows:

$$\Gamma = \sum_{\forall s \in S, \forall i \in F_s, \forall j \in \Psi^s_i} 1 \quad (1)$$

It shall be noted that in all the followings, $y_j$ represents the authorized nodes that that can be susceptible to host a given NF. We define the following variables: $\forall v \in V$:

$$\rho^{NODE}_v = \begin{cases} 1 & \text{If the virtual or physical node } v \text{ should be used} \\ 0 & \text{Otherwise} \end{cases} \quad (2)$$

We also define the following variables:
$\forall s \in S, \forall i \in F_s, \forall j \in \Psi^s_i \forall v \in y_j$:

$$Y_{s,i,j,v} = \begin{cases} 1 & \text{If the NF } j \text{ is running on top of node } v \\ 0 \end{cases} \quad (3)$$

In the Objective Function 4, we aim to minimize the number of nodes hosting the NFs that constitute different network slices.

$$\min \sum_{v \in V} \rho^{NODE}_v \quad (4)$$

Meanwhile, as discussed below, the constraints will be divided into five parts: the placement constraints, the resources constraints, the links arrangements constraints, the latency aware constraints and the bandwidth aware constraints.

### A. Placement Constraints

In this subsection, constraints related to the placement will be introduced. Indeed, Constraint 5 ensures that each NF should run on top of only one node;

$\forall s \in S, \forall i \in F_s, \forall j \in \Psi^s_i$:

$$\sum_{v \in y_j} Y_{s,i,j,v} = 1 \quad (5)$$

Constraint 6 shows that if a given NF is running on top of a node, this node must be used;

$\forall s \in S, \forall i \in F_s, \forall j \in \Psi^s_i \forall v \in y_j$:

$$\rho^{NODE}_v \geq Y_{s,i,j,v} \quad (6)$$

### B. Resources Constraints

In this subsection, constraints related to the resources will be discussed. We denote by $\mathcal{R}$ the set of resource types, such as CPU, RAM, Storage and so on. From the Slice Descriptor presented in Fig. 1, we can get the required resources for each NF.

$\forall s \in S, \forall i \in F_s, \forall j \in \Psi^s_i, \forall r \in \mathcal{R}$: where $R_{s,i,j}(r)$ denotes the required resource $r$ of NF $j$. We denote by $\pi^r_v$ the available amount of resources $r \in \mathcal{R}$ in node $v \in V$. Constraint 7 ensures that the number of resources is respected. Each NF request should not exceed the available resources in any given node deployed to serve network slices.

$v \in V, \forall r \in \mathcal{R}$:

$$\sum_{s \in S, i \in F_s, j \in \Psi^s_i} R_{s,i,j}(r) \times Y_{s,i,j,v} \leq \pi^r_v \quad (7)$$

### C. Links Arrangement Constraints

This subsection introduces all variables and constraints in relation to the links arrangement. From constraints 9 to 12, we define the variables that have a relationship with the Slice-physical layer. In the following, we assume that a node can host multiple NFs.

$$\forall s \in S, \forall i \in F_s, \forall j \in \{\Psi_i^s - \Psi_{i,1}^s\}, \forall u \in V, \forall v \in V:$$

$$Z_{s,i,u,v}^{j-1,j} = \begin{cases} 1 & \text{If the traffic between } j \text{ and } j-1 \text{ passes through the link } (u,v) \\ 0 & \text{Otherwise} \end{cases} \quad (8)$$

We have Constraint 9 that ensures the presence of a link between each two consecutive NFs.

$$\forall s \in S, \forall i \in F_s, \forall j \in \{\Psi_i^s - \Psi_{i,1}^s\}:$$

$$\sum_{u \in V} \sum_{v \in V} Z_{s,i,u,v}^{j-1,j} = 1 \quad (9)$$

We also have the following inequalities in 10, 11 and 12 which guarantee that if there is a link between $NF_{j-1}$ and $NF_j$, the node $u$ and $v$ are hosting respectively $NF_{j-1}$ and $NF_j$, deployed in the network:

$$\forall s \in S, \forall i \in F_s, \forall j \in \{\Psi_i^s - \Psi_{i,1}^s\}, \forall u \in V, \forall v \in V:$$

$$Z_{s,i,u,v}^{j-1,j} \leq Y_{s,i,j-1,u} \quad (10)$$

$$Z_{s,i,u,v}^{j-1,j} \leq Y_{s,i,j,v} \quad (11)$$

$$Z_{s,i,u,v}^{j-1,j} \geq Y_{s,i,j-1,u} + Y_{s,i,j,v} - 1 \quad (12)$$

### D. Latency Aware Constraints

The constraints considered in this subsection guarantee that the links have enough end-to-end latency for ensuring a good system functionality. In Constraints 13 to 18, the latency of the Slice-physical layer mapping is ensured. We define the following variables.

$\varphi_{j-1,j}^L$ is a real variable that shows the maximum delay between $NF_{j-1}$ and $NF_j$ in SFC $i$ in slice $s$. Constraint 13 ensures that the desired end-to-end latency is maintained in the slice layer.

$$\forall s \in S, \forall i \in F_s:$$

$$\sum_{j \in \{\Psi_i^s - \Psi_{i,1}^s\}} \varphi_{j-1,j}^L \leq l^{fi} \quad (13)$$

We also define the following. $\forall u \in V, \forall v \in V, \Phi_{u,v}^L$ represents the maximum delay between nodes $u$ and $v$. Constraint 14 ensures that if the communication between $NF_{j-1}$ and $NF_j$ uses the link $u, v$, then the delay between $node_u$ and $node_v$ must be longer than the delay between $NF_{j-1}$ and $NF_j$.

$$\forall u \in V, \forall v \in V, \forall s \in S, \forall i \in F_s, \forall j \in \{\Psi_i^s - \Psi_{i,1}^s\}:$$

$$\Phi_{u,v}^L \leq \varphi_{j-1,j}^L \times Z_{s,i,u,v}^{j-1,j} \quad (14)$$

However, inequality (14) is not linear. To make the optimization problem linear, we introduce the following variables and constraints. Firstly, we define the following real variables:
$\forall u \in V, \forall v \in V, \forall s \in S, \forall i \in F_s, \forall j \in \{\Psi_i^s - \Psi_{i,1}^s\}:$
$\varphi_{s,i,j}^{L,u,v}$ with M as a big number ($M \approx \infty$) and $\varphi_{s,i,j}^{L,u,v} = \varphi_{j-1,j}^L$ iff $Z_{s,i,u,v} = 1$, otherwise $\varphi_{s,i,j}^{L,u,v} = 0$, we add the following constraints:

$$\forall u \in V, \forall v \in V, \forall s \in S, \forall i \in F_s, \forall j \in \{\Psi_i^s - \Psi_{i,1}^s\}:$$

$$\varphi_{s,i,j}^{L,u,v} \leq \varphi_{j-1,j}^L + (1 - Z_{s,i,u,v}^{j-1,j}) \times M \quad (15)$$

$$\varphi_{j-1,j}^L \leq \varphi_{s,i,j}^{L,u,v} + (1 - Z_{s,i,u,v}^{j-1,j}) \times M \quad (16)$$

$$\varphi_{s,i,j}^{L,u,v} \geq (1 - Z_{s,i,u,v}^{j-1,j}) \times M \quad (17)$$

$$\Phi_{u,v}^L \leq \varphi_{s,i,j}^{L,u,v} \quad (18)$$

### E. Bandwidth Aware Constraints

The following constraints guarantee that each link has enough bandwidth for ensuring system functionality. Constraint 19 to 24 show that the bandwidth of the Slice-physical layer mapping is ensured. We define $\varphi_j^W$ as a real variable that shows the minimum bandwidth between $NF_{j-1}$ and $NF_j$ in SFC $i$ in slice $s$. Constraint 19 ensures that the required bandwidth is respected between each two successive NFs.

$$\forall s \in S, \forall i \in F_s, \forall j \in \{\Psi_i^s - \Psi_{i,1}^s\}:$$

$$\varphi_{j-1,j}^W \geq w^{fi} \quad (19)$$

Constraint 20 ensures that the bandwidth between $node_u$ and $node_v$ equals the sum of all the bandwidths used by different NFs. $\Phi_{u,v}^W$ represents the minimum bandwidth between the nodes $u$ and $v$.

$$\forall u, v \in V:$$

$$\Phi_{u,v}^B \geq \sum_{s \in S, i \in F_s, j \in \{\Psi_i^s\}} \varphi_{j-1}^B \times Z_{s,i,u,v}^{j-1,j} \quad (20)$$

However, inequality (20) is nonlinear. To make the optimization problem linear, we introduce the following variables and constraints.

Firstly, we add the following real variables: $\forall u, v \in V, \forall s \in S, \forall i \in F_s, \forall j \in \{\Psi_i^s - \Psi_{i,1}^s\}: \varphi_{s,i,j}^{W,u,v}$
With M as a big number ($M \approx \infty$) and $\varphi_{s,i,j}^{B,u,v} = \varphi_{j-1,j}^B$ iff $Z_{s,i,u,v} = 1$, otherwise $\varphi_{s,i,j}^{B,u,v} = 0$, we add the following constraints:

$$\forall u, v \in V, \forall s \in S, \forall i \in F_s, \forall j \in \{\Psi_i^s - \Psi_{i,1}^s\}:$$

$$\varphi_{s,i,j}^{B,u,v} \leq \varphi_{j-1,j}^B + (1 - Z_{s,i,u,v}^{j-1,j}) \times M \quad (21)$$

$$\varphi_{j-1,j}^B \leq \varphi_{s,i,j}^{B,u,v} + (1 - Z_{s,i,u,v}^{j-1,j}) \times M \quad (22)$$

$$\varphi_{s,i,j}^{B,u,v} \leq Z_{s,i,u,v}^{j-1,j} \times M \quad (23)$$

$$\forall u \in V, \forall v \in V:$$

$$\Phi_{u,v}^B \geq \sum_{s \in S, i \in F_s, j \in \{\Psi_i^s - \Psi_{i,1}^s\}} \varphi_{s,i,j}^{B,u,v} \quad (24)$$

Fig. 2 presents an example that illustrates the operations of our proposed solution. In this figure, the proposed architecture, which consists of six data-centers named from "A" to "I" in the underlying layer. Also note that for the sake of clarity, the bandwidth and the latency will be both represented by W in the underlying layer. We also omitted the **5G-VM**, the **SID** and the **SO** from Fig. 2. They are responsible for the creation of different kinds of network slices after receiving requests from end users.

Let assume that we have a connected car management

$$\varphi_{s,i,j}^{L,u,v} \leq \varphi^L + (\mathbf{1} - \mathbf{Z}^{j-1,j}) \times \mathbf{M} \qquad (15)$$

scenario. In this scenario, and for safety reasons, we need at

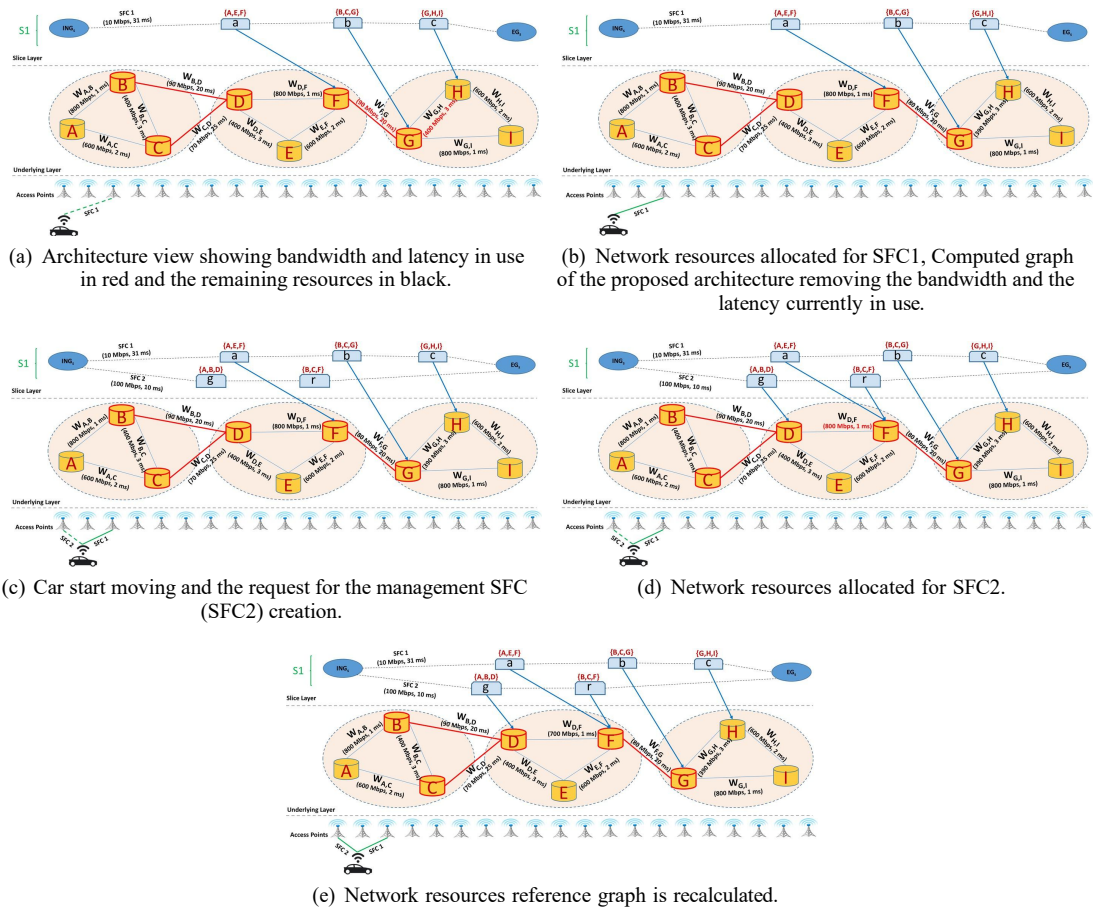

(a) Architecture view showing bandwidth and latency in use in red and the remaining resources in black.

(b) Network resources allocated for SFC1, Computed graph of the proposed architecture removing the bandwidth and the latency currently in use.

(c) Car start moving and the request for the management SFC (SFC2) creation.

(d) Network resources allocated for SFC2.

(e) Network resources reference graph is recalculated.

Fig. 2: Example of our proposed solution for cost optimal network slicing deployment.

least one mandatory network slice containing a SFC dedicated for the monitoring and measurement tasks in case of mobility. In Fig. 2(a), we assume that a passenger on board a connected car requests a video streaming service that requires $10Mbps$ of bandwidth and a latency of 31ms. By collecting the request, the **5G-VM**, the **SID** and the **SO** will coordinate and create a slice dubbed "S1" in the Slice layer, "S1" contains 3 NFs "a", "b" and "c" deployed in nodes F, G, and H, respectively. Fig. 2(a) shows the bandwidth and latency resources partially in use as highlighted by the red numbers between nodes F, G, and H in the underlying layer. Based on this topology, we update our reference graph $G$ by removing all used resources as depicted in Fig. 2(b). In Fig. 2(c), the connected car starts moving and as explained earlier, for safety reasons, a second SFC is needed to handle the monitoring tasks and the measurements. Therefore, the connected car requests the creation of this second SFC. However, that second SFC is resource consuming and needs at least 100Mbps of bandwidth and 10ms of latency as it is a delay-sensitive service. In addition, $NF_g$ and $NF_r$ have some restrictions in terms of deployment on the underlying layer. For instance, $NF_r$ can only be deployed on nodes B, C and F; here we can clearly observe Constraints 5 and 6. Fig. 2(d) shows the bandwidth and latency resources partially in use, being highlighted using the red color between nodes D and F in the underlying layer. Fig. 2(e) represents the graphs $G$ in the underlying layer after the re-computation. It shall be noticed that all constraints related to the link arrangement, latency and bandwidth are used in each request by our proposed solution.

Algorithm 1 summarizes these constraints. For each new request for slice creation, the distribution of NFs that form the slices has to be recomputed. This algorithm is triggered by either the arrival of a new request or an updated one: whether it regards the arrival of a new user or the mobility of an existing one. The input parameters of the proposed solution are the graph $G$ that represents the underlying layer and the requests' specifications (i.e., authorized nodes, bandwidth, end-to-end latency, etc). Every time the algorithm returns a new configuration, the main control loop is re-executed waiting for an update in our proposed network topology.

### F. Final model

After introducing all constraints and their respective transformations, the final model to optimize is:

**Algorithm 1** FQER Algorithm.

**Require:**
  $G$: Network graph.
  $Q$: List of Requests' specification (resources, authorized hosts, bandwidth, end-to-end latency).
  **while** true **do**
    **if** $q$ in $Q$ == (*new or updated*) **then**
      $Optimization_{FQER}(G, Q)$
    **end if**
  **end while**

S.t $\min \sum_{v \in \{1...V\}} \rho_v^{NODE}$
  ; Constraints 5 – 7.
    Constraints 9 – 13.
    Constraints 15 – 19.
  ; Constraints 21 – 24.

## V. PERFORMANCE EVALUATION

We created a simulator using Python, based on the Gurobi optimization solver to implement our optimization model and solve the previously formulated problem. The underlying layer's components (i.e nodes, edges, CPU, RAM, Disk), network slices, SFCs, NFs, and the NFs' resources requirements are randomly generated to simulate a more realistic environment. The resource demands of each NF, in terms of both bandwidth and latency between NFs, follow a discrete uniform distribution over the interval $[50, 100]$. We conducted our experiments on a multi-core server as described in Table I.

TABLE I: Hardware Configuration.

| Type | Configuration |
|---|---|
| CPU | Dual Intel Xeon E5-2680 v3 @ 2.5GHz |
| Memory | 256GB |
| Linux | Ubuntu 16.04 |
| Kernel | 4.4.0-72 |

We started the evaluation of our solution's behavior by varying the number of network slices, SFCs, and NFs. For each experiment, we operate 100 repetitions, changing the underlying layer's components deployment and compute the number of nodes' used as well as the computational times. Afterward, we present the mean and $95\%$ Confidence Interval of the number of nodes' used in the proposed architecture and the computational cost in seconds.

Fig. 3(a) presents the behavior of our solution when we vary the number of network slices over the physical network from 1 to 50, while keeping for each slice both the number of SFCs and NFs constant, i.e., running our simulations with 2 and 4, respectively. In Fig. 3(a), the left Y-axis represents the number of nodes used, the right Y-axis shows the required time in seconds to solve the optimization problem, and the X-axis portrays the number of network slices in the proposed architecture. Fig. 3(a) shows that the number of active nodes increases from 3 to 8 when we reach the 15th variation of network slices, then, from 15 to 20 network slices, we observe a relative stagnation between 8 and 9 activated nodes, while beyond 20 network slices, the mean number of nodes starts to stabilize. The mean number of nodes activated is $8.3798$ with a standard deviation of $1.6638$. Meanwhile, the right Y-axis shows that as the number of network slices increases, the computational cost increases linearly from $0.1417s$ to $11.9012s$. The linear regression parameters are $0.221$ and $0.628$, as $\alpha$ and $\beta$, respectively, for the solution cost of our set of experiments.

In Fig. 3(b), we vary SFCs number in the network from 2 to 20 with the length of SFCs set as 4. We kept the same representations as before for the Y-axis but for the X-axis we are considering the number of SFCs in our network instead of network slices. We observed that our results start to stabilize after 7 SFCs, with a mean number of activated nodes of $9.1779$ and a standard deviation of $0.4813$. Still, in Fig. 3(b), we noticed that the solution cost in seconds increases linearly with the number of activated nodes in the network: the linear regression of these samples was $0.443$ and $0.140$, as $\alpha$ and $\beta$, respectively. We must also state that our evaluation assumes at least one SFC for the management tasks, which causes our simulation to start from 2 SFCs.

In Fig. 3(c), we use 5 network slices and 2 SFCs in our evaluation while varying the number of NFs in the network. Different from the previous experiment, we can perceive an exponential growth in the computational time. The reason is that when varying the number of NFs from 2 to 20, we are actually rising the total number of NFs from 20 to 200 since this number is multiplied by the number of SFCs and slices in a fixed number of underlying layer with a high rate of variations.

The experiment in Fig. 3(d) aims for comparing small-scale underlying topologies against large-scale ones. The small- scale case is represented by the red color and the blue one where the number of nodes used and computational time is the results when we vary the number of network slices while fixing the number of SFCs, NFs and underlying nodes to 2, 4 and 12, respectively. Concerning the large-scale networks, the green color and the orange color illustrate the number of nodes used and computational time of the evaluation for the proposed solution when we vary the number of network slices in addition to the underlying nodes while fixing the number of SFCs and NFs to 2 and 4, respectively.

The first observation that we can draw from this figure is that the large-scale networks have slightly better performance than small-scale networks and that is in terms of the number of underlying nodes used regardless the number of network slices and the number of underlying nodes varied during the simulation. However, we clearly observe that the small- scale networks have better performance than the large ones in terms of execution time as the computational time grows up exceptionally compared to when we fix the number of nodes to 12, which conserves the linear trajectory. From this behavior, we can conclude that when we increase the number

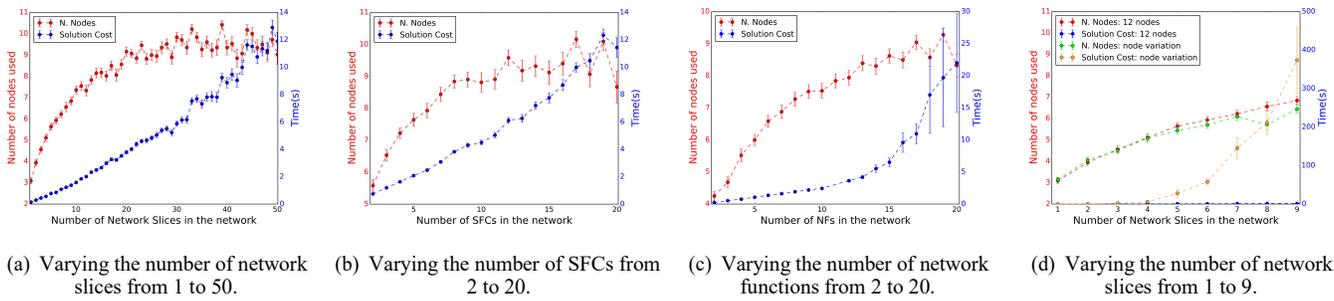

(a) Varying the number of network slices from 1 to 50.

(b) Varying the number of SFCs from 2 to 20.

(c) Varying the number of network functions from 2 to 20.

(d) Varying the number of network slices from 1 to 9.

Fig. 3: Performance evaluation results.

of underlying nodes, the number of links between those nodes increases due to the higher density of nodes' in our network which will raise the probability of deployment of the network slices causing by the same time the exponential growth in the computational time.

## VI. CONCLUSION

In this article, we introduced a novel approach for optimal deployment of network slices taking cost into account. We evaluated the proposed scheme using multiple network topologies, in particular comparing small-scale topologies against large-scale networks. Interesting results were obtained when varying both the number of underlying nodes and the number of network slices, as we observed that the computational time grows exceptionally compared to when we set the number of underlying nodes to a fix number, which conserves the linear trajectory. In future works, we will consider finding a solution to compute the underlying node distribution in a polynomial time for real-world deployment by proposing heuristic algorithms able to outperform the NP-hardness problem faced. This is in addition to investigating the mobility feature of Network Slices to fulfill mobile users' requirements.

## ACKNOWLEDGMENT

This work was supported in part by the Academy of Finland Project CSN under Grant No. 311654. The work was also supported in part by a direct funding from Nokia Bell Labs, Espoo, Finland.

## REFERENCES

[1] N. Alliance, "5G white paper," Tech. Rep., February 2015. [Online]. Available: https://www.ngmn.org/uploads/media/NGMN_5G_White_Paper_V1_0.pdf
[2] T. Taleb, K. Samdanis, B. Mada, H. Flinck, S. Dutta, and D. Sabella, "On multi-access edge computing: A survey of the emerging 5g network edge cloud architecture and orchestration," *IEEE Communications Surveys Tutorials*, vol. 19, no. 3, pp. 1657–1681, thirdquarter 2017.
[3] M. Bagaa, T. Taleb, A. Laghrissi, A. Ksentini, and H. Flinck, "Coalitional Game for the Creation of Efficient Virtual Core Network Slices in 5G Mobile Systems," *IEEE Journal on Selected Areas in Communications*, vol. 36, no. 3, pp. 469–484, March 2018.
[4] I. Benkacem, T. Taleb, M. Bagaa, and H. Flinck, "Optimal VNFs Placement in CDN Slicing Over Multi-Cloud Environment," *IEEE Journal on Selected Areas in Communications*, vol. 36, no. 3, pp. 616–627, March 2018.
[5] A. Laghrissi, T. Taleb, and M. Bagaa, "Conformal Mapping for Optimal Network Slice Planning Based on Canonical Domains," *IEEE Journal on Selected Areas in Communications*, vol. 36, no. 3, pp. 519–528, March 2018.
[6] T. Taleb, M. Bagaa, and A. Ksentini, "User mobility-aware Virtual Network Function placement for Virtual 5G Network Infrastructure," in *2015 IEEE International Conference on Communications (ICC)*, June 2015, pp. 3879–3884.
[7] I. Afolabi, T. Taleb, K. Samdanis, A. Ksentini, and H. Flinck, "Network slicing; softwarization: A survey on principles, enabling technologies; solutions," *IEEE Communications Surveys Tutorials*, vol. PP, no. 99, pp. 1–1, 2018.
[8] J. Ordonez-Lucena, P. Ameigeiras, D. Lopez, J. J. Ramos-Munoz, J. Lorca, and J. Folgueira, "Network slicing for 5g with sdn/nfv: Concepts, architectures, and challenges," *IEEE Communications Magazine*, vol. 55, no. 5, pp. 80–87, May 2017.
[9] G. P. A. W. Group, "View on 5g architecture," Tech. Rep., July 2016. [Online]. Available: https://5g-ppp.eu/wp-content/uploads/2014/02/5G-PPP-5G-Architecture-WP-For-public-consultation.pdf
[10] H. Hantouti, N. Benamar, T. Taleb, and A. Laghrissi, "Traffic Steering for Service Function Chaining," *IEEE Communications Surveys Tutorials*, pp. 1–1, 2018.
[11] A. M. Medhat, T. Taleb, A. Elmangoush, G. A. Carella, S. Covaci, and T. Magedanz, "Service Function Chaining in Next Generation Networks: State of the Art and Research Challenges," *IEEE Communications Magazine*, vol. 55, no. 2, pp. 216–223, February 2017.
[12] Z. Kotulski, T. Nowak, M. Sepczuk, M. Tunia, R. Artych, K. Bocianiak, T. Osko, and J. P. Wary, "On end-to-end approach for slice isolation in 5g networks. fundamental challenges," in *2017 Federated Conference on Computer Science and Information Systems (FedCSIS)*, Sept 2017, pp. 783–792.
[13] H. Moens and F. D. Turck, "Vnf-p: A model for efficient placement of virtualized network functions," in *10th International Conference on Network and Service Management (CNSM) and Workshop*, Nov 2014, pp. 418–423.
[14] H. Ko, D. Suh, H. Baek, S. Pack, and J. Kwak, "Optimal placement of service function in service function chaining," in *2016 Eighth International Conference on Ubiquitous and Future Networks (ICUFN)*, July 2016, pp. 102–105.
[15] X. Song, X. Zhang, S. Yu, S. Jiao, and Z. Xu, "Resource-efficient virtual network function placement in operator networks," in *GLOBECOM 2017 - 2017 IEEE Global Communications Conference*, Dec 2017, pp. 1–7.
[16] S. Jiao, X. Zhang, S. Yu, X. Song, and Z. Xu, "Joint virtual network function selection and traffic steering in telecom networks," in *GLOBECOM 2017 - 2017 IEEE Global Communications Conference*, Dec 2017, pp. 1–7.
[17] H. Li, L. Wang, X. Wen, Z. Lu, and L. Ma, "Constructing service function chain test database: An optimal modeling approach for coordinated resource allocation," *IEEE Access*, vol. PP, no. 99, pp. 1–1, 2017.
[18] T. Wen, H. Yu, G. Sun, and L. Liu, "Network function consolidation in service function chaining orchestration," in *2016 IEEE International Conference on Communications (ICC)*, May 2016, pp. 1–6.